\documentstyle[12pt,aasms4]{article}

\slugcomment{To appear in The Astrophysical Journal Letters}

\lefthead{Marani et al.}
\righthead{Lensed GRBs as Probes of Dark Compact Objects}

\begin{document}

%\title{Gravitationally Lensing, Dark Matter, and Gamma-Ray Bursts}
\title{Gravitationally Lensed Gamma-Ray Bursts as Probes of \\
Dark Compact Objects}

\author{G.F. Marani\altaffilmark{1,2}, R.J. Nemiroff\altaffilmark{3},
J.P. Norris\altaffilmark{2}, K.Hurley\altaffilmark{4}, and J.T.
Bonnell\altaffilmark{2,5}}

\altaffiltext{1}{CEOSR George Mason University, Fairfax, VA 22030;
gmarani@science.gmu.edu}
\altaffiltext{2}{NASA Goddard Space
Flight Center, Greenbelt, MD 20771}
 \altaffiltext{3}{Department of Physics, Michigan
Technological University, Houghton, MI 49931}
\altaffiltext{4}{University of California at Berkeley, Space
Sciences Laboratory, Berkeley, CA 94720-7450}
 \altaffiltext{5}{Universities Space Research Association}

%========================================================
\begin{abstract}
If dark matter in the form of compact objects comprises a large
fraction of the mass of the universe, then gravitational lensing
effects on gamma-ray bursts are expected. We utilize BATSE
and Ulysses data to search for lenses of different mass ranges,
which cause lensing in the milli, pico, and femto regimes. Null
results are used to set weak limits on the cosmological abundance
of compact objects in mass ranges from 10$^{-16}$ to 10$^{-9}$
$M_{\odot} $.  A stronger limit is found for a much discussed
$\Omega = 0.15$ universe dominated by black holes of masses $\sim
10^{6.5} M_{\odot}$, which is ruled out at the $\sim$ 90\%
 confidence level.

\end{abstract}

%========================================================

%========================================================
\section{Introduction}

The possibility that a cosmological abundance of dark compact
objects (COs) in the universe could be revealed by gravitational
lensing of more distant sources was first suggested by Press \&
Gunn (1973). The detection of lensed images from any source
provides an estimation of the fraction of the universe composed of
COs, and therefore the CO density. Non-detections offer a viable
way to set constraints on the possible values of the CO mass,
$M_{\scriptscriptstyle CO}$, and density,
$\Omega_{\scriptscriptstyle CO}$. Depending on the angular
separation of the lensed images, the gravitational lensing effect
is classified into {\it milli-, micro-, pico-, or femto-lensing}.

There is now overwhelming evidence that the majority of gamma-ray
burst (GRB) sources lie at cosmological distances
(\cite{970508,971214,980703}). Their short duration and the
transparency of the Universe in gamma rays make GRBs ideal probes
of dark matter in the form of COs over the mass range $10^{-16}$
to $10^{12} M_{\odot}$. The probability of lensing a GRB depends
on CO density (e.g. \cite{Nem.vol,mao92}), but it is estimated to
be $< 0.1$\% (\cite{Blaes}).

Multiple images of the same burst cannot be angularly resolved by
present gamma-ray detectors, but they are easily temporally
resolved (\cite{pac2}). Thus, rather than spatially identifying
multiple images of lensed quasars, the lensed GRBs must be found
by searching such events with identical temporal and spectral
signatures. Marani et al. (1998) have compared the lightcurves of
1235 BATSE bursts, searching for gravitationally lensed GRBs
caused by foreground galaxies, with negative results. The expected
time delay between two detectable images can be written as (e.g.
\cite{Krauss})
\begin{equation} \Delta t =
{{ R_s (1+z_L)} \over {c}} \left( {{f-1} \over {\sqrt{f}}} - \ln f
\right),
\end{equation} where $f$ is the
brightness ratio of the two images also known as the dynamic
range, $z_L$ is the lens redshift, $R_s \sim 3$ km
$(M_{\scriptscriptstyle CO}/M_{\odot})$ is the Schwarzchild
radius, and $c$ is the speed of light. Mao (1992) found that for a
point mass lens, the median value of the time delay is $\Delta t
\sim 50$ s $\times (M_{\scriptscriptstyle CO}/10^6 M_{\odot})$.

In this paper, we present results of searches for different types
of gravitational lensing of GRBs, looking for any indication of
the presence of dark matter in the form of compact objects. In all
the cases, a monoluminous and non-evolving GRB source population
is assumed. In \S 2, we describe the lensing types, search
algorithms performed using BATSE data (\cite{Meegan}), and present
results. We summarize and discuss these results in \S 3.

%========================================================
\section{Searching for Dark Compact Objects}

%------------------------
\subsection{Millilensing}

In the case of millilensing, the COs have masses between $10^{6}$
and $10^{9}$ $M_{\odot}$. The time delay between the images may
vary from few seconds to $\sim 10^4$ seconds and their average
angular separation  is $\sim 10^{-3}$ arcseconds. Note that
$M_{\scriptscriptstyle CO}= 10^{6.5} M_{\odot}$  is the Jean mass
at recombination in many early universe scenarios and,
consequently, a probable mass scale for dark matter. Gnedin \&
Ostriker (1992) proposed an open universe model with $\Omega_M =
\Omega_{\scriptscriptstyle CO} = 0.15$, filled with black holes of
mass $ \sim 10^{6.5} M_{\odot}$, which would dominate the dark
matter galactic halos.

Nemiroff et al. (1993) searched the first 44 BATSE bursts for
lensed images that came shortly after the main event by
autocorrelating each burst, and no lens candidates were found.
This null result was interpreted by using the detection volume
formalism of Nemiroff (1989), and by modeling the brightness
distribution to estimate GRB redshifts (\cite{Thulsi}). A standard
(1,0) universe composed to closure density with objects of mass
$10^{6.5}$ - $10^{8.1} M_{\odot}$ was marginally ruled out
(\cite{Nem.mega}).

The data from Table I of Nemiroff et al. (1993) and their null
results were used to set constraints on the CO mass and density
for the Gnedin \& Ostriker universe. Mao (1993) estimated that the
lensing probability for this universe is $\sim$ 5\%. Following
Nemiroff et al. (1993), for each burst we compute a detection
volume between the observer and a burst at $z_S$, within which the
lens with $M_{\scriptscriptstyle CO}$ must fall to create a
detectable image. Then, all 44 volumes are added and the expected
number of detectable lenses in our sample is obtained. For a given
($\Omega_{\scriptscriptstyle CO} , M_{\scriptscriptstyle CO}$)
pair, if the expected number is significantly greater than the
null result obtained, we rule out such CO parameters. For a more
detailed explanation of the method, see Nemiroff et al. (1993) and
Marani (1998b).

Figure~\ref{fig1} summarizes the results we  obtain for the Gnedin
\& Ostriker universe. The figure shows the number of images
expected for the 44 GRBs and for different
($\Omega_{\scriptscriptstyle CO}, M_{\scriptscriptstyle CO}$)
parameters. The expected number of images for the Gnedin \&
Ostriker universe is $\sim$ 2.5 - 3, in good agreement with Mao's
estimations (1993): 2.2 lenses from 44 bursts  were expected.
Since no images were found, the universe with
\begin{equation}
M_{CO} \sim  10^{6.5} M_{\odot}, \ \ \ \ \  \
\Omega_M = \Omega_{CO} = 0.15
\end{equation}
can be excluded at a confidence level of $\sim$ 90.0\%.

%-----------------------
\subsection{Picolensing}

Picolensing refers to lensing mass range of $\sim
10^{-12}$-$10^{-7} M_{\odot}$. The expected angular separation is
$\sim 10^{-12}$ arcseconds and the average time delay is $\sim
10^{-18}$ seconds. The main lens candidates are primordial black
holes, molecular clouds, and small planets. Currently, few
arguments favor the hypothesis that extremely low mass objects
compose a significant fraction of the dark matter. Nevertheless,
if such objects were to exist, GRB lensing is an efficient method
for their detection (\cite{Nem.pico}).

For these low mass COs, the corresponding radius of the Einstein
ring projected onto our solar system, $\tilde{R}_E$, will be at
most 1 AU, $\tilde {R}_E = \sqrt{ ( 2 R_s D_{\scriptscriptstyle
OL} D_{\scriptscriptstyle OS} )/ D_{\scriptscriptstyle LS} }$. Two
spacecraft separated by 1 AU or more might see the same GRB event
with very different amplification since, if the GRB were to lie
within the $\tilde{R}_E$ of the lens for one spacecraft, it would
lie outside $\tilde{R}_E$ for the other spacecraft. The $i^{th}$
spacecraft observes the GRB with total amplification $A_i =
(\tilde{\beta}^2 + 2)/(\tilde{\beta} \sqrt{\tilde{\beta}^2 + 4})$,
where $\tilde{\beta}= \beta/\beta_E$, with $\beta$ the angular
separation between source and lens, and $\beta_E$ the angular
radius of the Einstein ring. There are currently two GRB detectors
on orbit separated by a distance of at least 1 AU: BATSE and
Ulysses GRB (\cite{hurley.uly}). If BATSE detects a strong burst
while Ulysses determines it is a weak burst, the difference in the
amplitude may be attributable to a gravitational lensing event due
to masses in the range  $\sim 10^{-12}$- $10^{-7} M_{\odot}$.

We have compared the peak counts of 60 bright bursts which were
triggered events in BATSE and Ulysses data
(\cite{Maranigl2,Marani.Thesis}). The energy range of Ulysses GRB
detector is $\sim$ 20-150 keV. The Ulysses triggered data have
temporal resolutions of 8- and 32-ms, but they were rebinned into
64-ms count bins. For BATSE data, we have used 64-ms concatenated
ASCII data summed over energy channels 1 and 2, corresponding to
the energy range $\sim$ 20-100 keV.

Once the data were background-subtracted and rebinned, the peak
counts and the BATSE-to-Ulysses amplitude ratios,
$dA=C_{p,\scriptscriptstyle BATSE}/ C_{p,\scriptscriptstyle
Ulysses}$, were calculated. No evident discrepancy in the
amplitude ratios has been found. Figure~\ref{fig2} shows BATSE
peak counts versus Ulysses peak counts. The error bars are
obtained assuming Poisson counts. The median value for the ratio,
after normalization, is $dA =1$, and the 2-$\sigma$ upper and
lower confidence levels are 2 and 0.68, respectively. A
real difference in amplitude should be observable in all three
timescales. We also compared peak counts for 256-ms and 1024-ms
time resolutions, although no significant excess was found.

These null results are used to impose limits on the density and
mass of COs, again using the detection volume formalism
(\cite{Nem.vol}). In the case of two observers, where a positive
detection requires an amplitude difference between them, the
volume within which a point-like lens must fall to create a
discernible magnification is much more complicated than in the
millilensing case, and no analytical expression is available. To
find the expected number of lenses, we implemented a Monte Carlo
simulation (\cite{Marani.Thesis}): For each burst in the sample,
we randomly draw several lens positions on the lens plane -- fixed
at $z_L$. We then compute the amplifications each observer would
see if the lens is located in that particular position, take the
ratio of those amplifications, and check if that ratio is above or
below the 2-$\sigma$ C.L. Finally, we place the lens plane at
another $z_L$, and re-compute the steps described above. When the
lens plane reaches the source plane at $z_S$, we obtain the
detection volume $V_{det}$ and the expected number of detectable
lenses for that particular burst, given by
$N_{det}=n_{\scriptscriptstyle CO} V_{det}$. Here,
$n_{\scriptscriptstyle CO}=3 H_o^2(1+z_L)^3
\Omega_{\scriptscriptstyle CO}/(4 \pi R_s c^2)$ is the CO comoving
number density. We then add all volumes of all bursts and repeat
the calculations for different CO masses and densities.

Figure~\ref{fig3} shows the results we obtained with this simulation, assuming
an Einstein-de Sitter universe.
Less than $\sim 0.9$ detectable images are expected in our
sample. Thus, the mass and density ranges of
\begin{equation}
M_{\scriptscriptstyle CO} \sim 10^{-12.5} \ -  \ 10^{-9.0} M_{\odot}, \ \ \ \ \
\Omega_{\scriptscriptstyle CO} \ge 0.9,
\end{equation}
may be ruled out only at $\sim$ 1-$\sigma$ statistical significance.

%-----------------------
\subsection{Femtolensing}

The femtolensing mass range is $M_{\scriptscriptstyle CO}$
$\sim$$10^{-16}$-$10^{-13} M_{\odot}$.
The expected image separation is $\sim 10^{-15}$
arcseconds, and the average time delay between the images is $\sim
10^{-24}$ seconds. Snowballs, primordial black holes, and comets are
some examples of these COs.

These extremely low mass lenses would produce interference
fringes in the GRB energy spectrum. These features are
evenly separated in energy space and might be confused with cyclotron
absorption lines (\cite{gould92}). Spectral lines between 10 and 100 keV
were seen in missions like {\it KONUS} (\cite{Mazets}),
{\it HEAO 1} (\cite{Hueter}), and {\it Ginga} (\cite{Fen1,Murakami}),
and indeed have been attributed to cyclotron lines due to magnetic
fields $B \sim 10^{12}$ G. However, these interpretations are still controversial.
Briggs (1996) pointed out that many spectral line analyses performed in
these GRB data sets used simplistic continuum models and incomplete statistical
techniques.

Several authors have analyzed BATSE spectra for spectral line
features like the absorption lines detected by {\it Ginga} at
$\sim$ 20 and $\sim$ 40 keV (see, for example,
\cite{Band2,Band3,Briggs2}). Briggs et al. (1998) have recently
performed a computerized search of 118 bright BATSE bursts. They
have found only 4 candidates with one possible absorption line,
but none of them with evenly spaced, multiple lines. Note that
Stanek et al. (1993) and Ulmer \& Goodman (1995) have argued that
the detectability of the femtolensing signature depends on the
source size. The more extended the source, the less pronounced the
spectral lines. Lines at higher energies are progressively weaker
for finite angular extent, since the source appears larger at
higher energies. However, the short duration of GRBs (the vast
majority lasting from milliseconds to a few tens of seconds) and
their cosmological distances suggest that GRBs can be considered
compact sources.

Assuming that GRB sources subtend sufficiently small angular sizes
to produce multiple, detectable lines in the spectra, the
non-detections of spectral lines in 118 BATSE bursts by Briggs et
al. (1998) can be used to impose lower limits on the mass and
density of COs. Nemiroff (1989) estimated that the probability of
lensing is $\sim \Omega_{\scriptscriptstyle CO}/8$ if the average
distance to the GRBs is $z_{GRB}\sim 1$, or $\sim
\Omega_{\scriptscriptstyle CO}/3$ if $z_{GRB} \sim 2$. Therefore,
we can rule out, at 95 \% confidence levels, universes composed of
COs with $M_{\scriptscriptstyle CO} \sim 10^{-16} - 10^{-13}
M_{\odot}$ and
\begin{equation}
  \Omega_{CO} \geq 0.2 \ , \mbox{ if}  \ z_{GRB} \sim 1; \
  \mbox{or} \ \ \Omega_{CO} \geq 0.1 \ , \mbox{ if}  \ z_{GRB} \sim 2.
\end{equation}
The above estimations assume an Einstein-de Sitter cosmology
and point mass lenses.

%========================================================
\section{Summary and Discussion}

The non-detections of lensing images in the current GRB data
have been used, for the first time, to set conservative
limits on the composition of dark compact object with masses
 between $10^{-16}$ and $10^{-7}$ $M_{\odot}$.
Different searches in radio, optical, and gamma ray data have been
conducted in the past to search for gravitationally lens-induced
images caused by masses between $10^{-6}$ and $10^{12}$
M$_{\odot}$ (\cite{Carr2}), but no search previous to the present
one was carried out for the pico- and femto-lensing cases. The
derived constraints on the density and mass parameters in
different mass ranges are shown in Figure~\ref{fig4}. The shaded
areas show the excluded values of $\Omega_{\scriptscriptstyle CO}$
and $M_{\scriptscriptstyle CO}$. Light gray shadings represent
very weak limits, whereas dark gray shadings are the imposed
limits at $\sim$ 2-$\sigma$ confidence level. Three additional
regions are shown in Figure~\ref{fig4}. The cross-hatched area
corresponds roughly to the range of values required to explain the
MACHO and EROS microlensing results (\cite{Carr2}). In the
millilensing region, a universe proposed by Carr \& Rees (1984)
with ($\Omega_M,\Omega_{\Lambda}) = (1,0)$ and
$\Omega_{\scriptscriptstyle CO} \sim 0.15, M_{\scriptscriptstyle
CO} \sim 10^6 M_{\odot}$ is shown as a hatched, very small area.
The constraints previously found by Nemiroff et al. (1993) are
also illustrated. Using gravitational lensing data from the
millilensing search done by these authors and the detection volume
formalism (\cite{Nem.vol}), we were able to reject a popular
universe proposed by Gnedin \& Ostriker (1992)
 with $M_{CO} \sim  10^{6.5} M_{\odot}$
and $\Omega_M = \Omega_{\scriptscriptstyle CO} = 0.15$, at a
confidence level of $\sim$ 90\%.

No search for microlensing events ($\Delta t \sim$ 5-10ms) in
BATSE archival data was conducted in the present work. However,
Nemiroff et al. (1998) have argued that such short events might
not be detected by BATSE due to its threshold setting for 64-ms
timescale.

The millilensing and picolensing limits are strong functions of
the GRB redshift distribution, which is derived by modeling the
observed brightness distribution. This model assumes no rate
density evolution, standard candle, and a simple power-law
spectrum, $dN/dE \propto E^{-1}$. Several authors have recently
suggested that the GRB rate density is proportional to the star
formation rate in the universe (\cite{Totani,Wijers}) and that the
intrinsic peak luminosities are described by a power-law
distribution (\cite{Kommers,Krumholz,mao98}). In both cases,
larger redshifts are expected. Therefore, the probability of
lensing by COs would increase and the limits imposed on
$\Omega_{\scriptscriptstyle CO}$ and $M_{\scriptscriptstyle CO}$
would be stronger.

%========================================================
\acknowledgments

This research was supported by grants from NASA.  Ulysses data
reduction is supported by JPL Contract 958056, and by NASA Grant
NAG 5-1560. GFM acknowledges support from a National Research
Council Research Associateship while at NASA GSFC. RJN
acknowledges additional support from the NSF.

\clearpage
%========================================================

\clearpage
%========================================================
\figcaption[fig1.ps]{Expected number of detectable lensing images
 for the given ($\Omega_{\scriptscriptstyle CO}$, $M_{\scriptscriptstyle CO}$)
values, in the ($\Omega_M$,$\Omega_{\Lambda}$)=(0.15,0)
universe proposed by Gnedin \& Ostriker (1992). The first available 44
BATSE bursts are incorporated in this plot.
\label{fig1}}

\figcaption[fig2.ps]{Ulysses GRB peak counts versus BATSE peak
counts for 60 bursts which were triggered events in both
detectors. The solid line represents the median value for the
64-ms peak count ratio, whereas the dashed lines enclose the
2-$\sigma$ confidence level region centered in the median value.
\label{fig2}}

\figcaption[fig3.ps]{Expected number of detectable picolensing images
for the given ($\Omega_{\scriptscriptstyle CO}$,$M_{\scriptscriptstyle CO}$)
values, incorporating 60 bursts that simultaneously triggered Ulysses GRB
and BATSE.
\label{fig3}}

\figcaption[fig4.ps]{Constraints on the density and mass of individual
compact objects, derived from the null results of gravitational
lensing searches in BATSE data.
\label{fig4}}

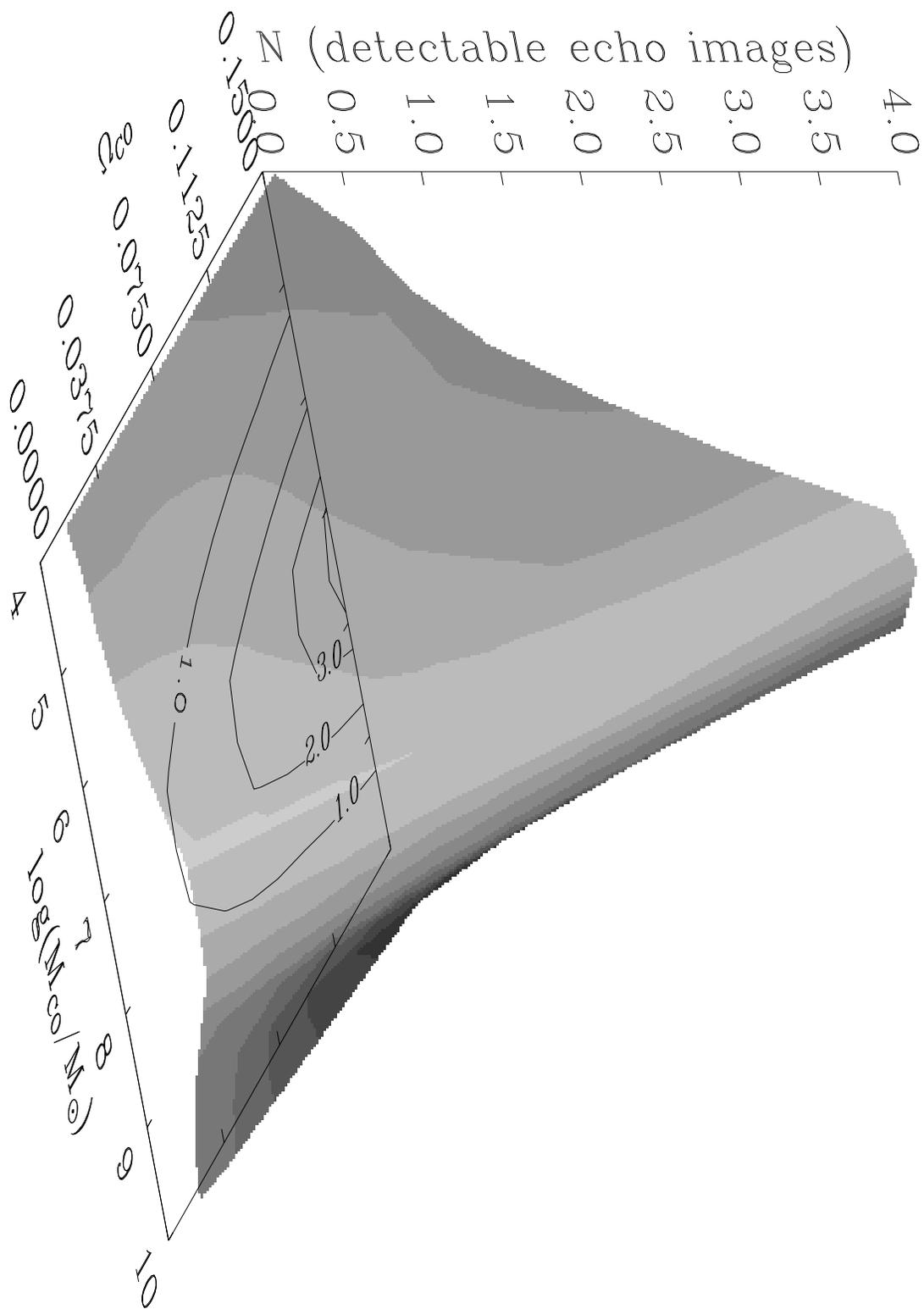
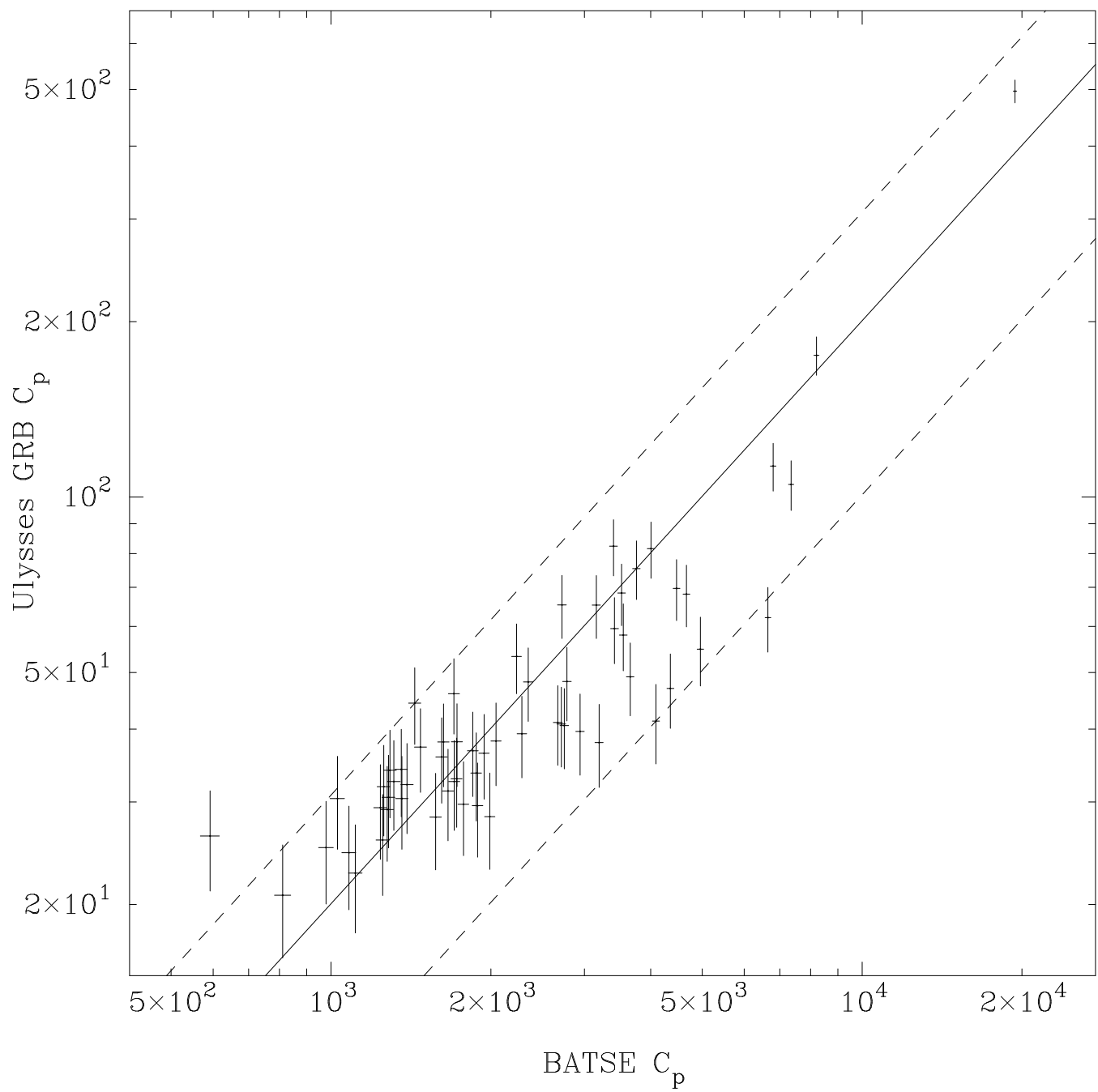
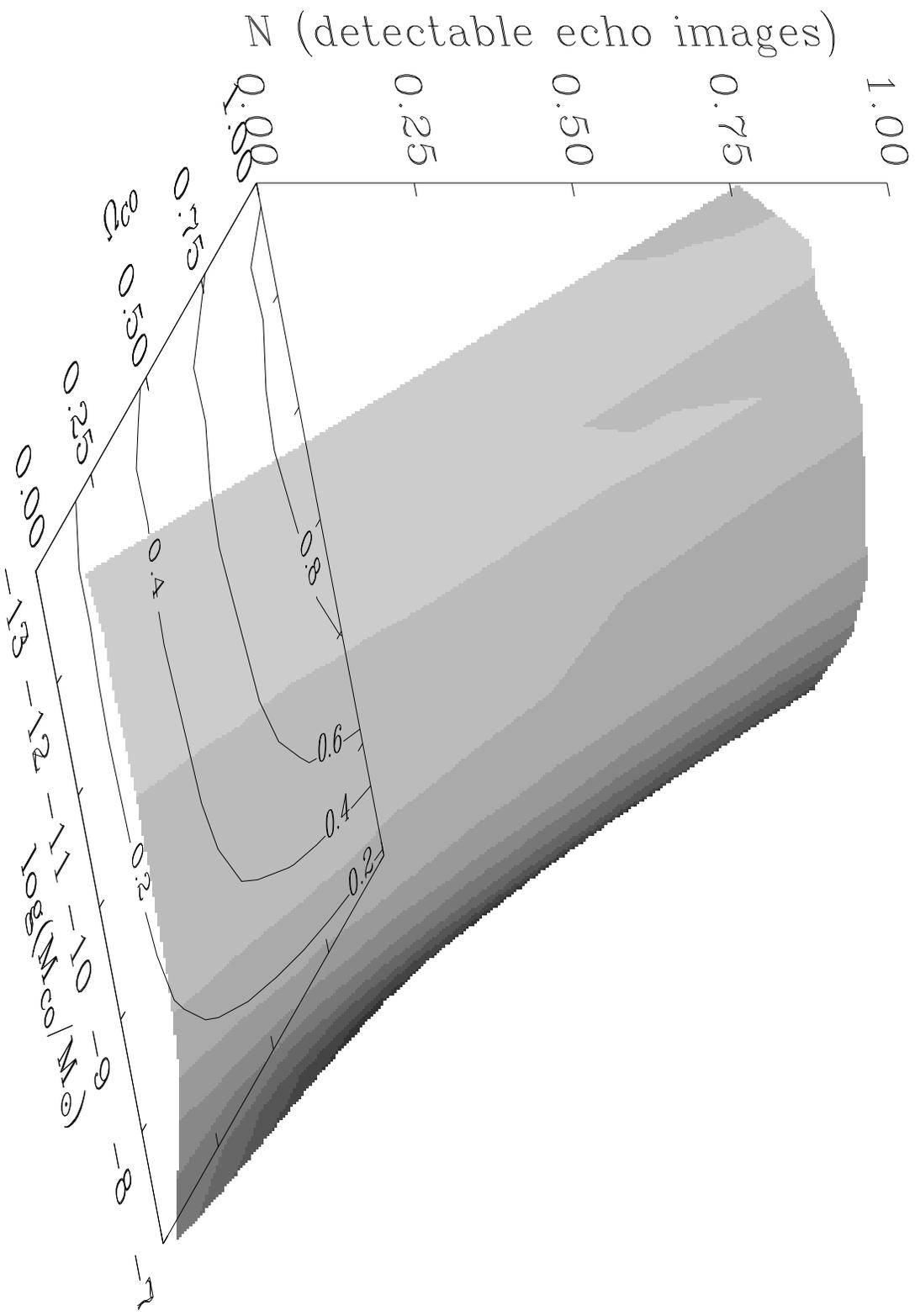
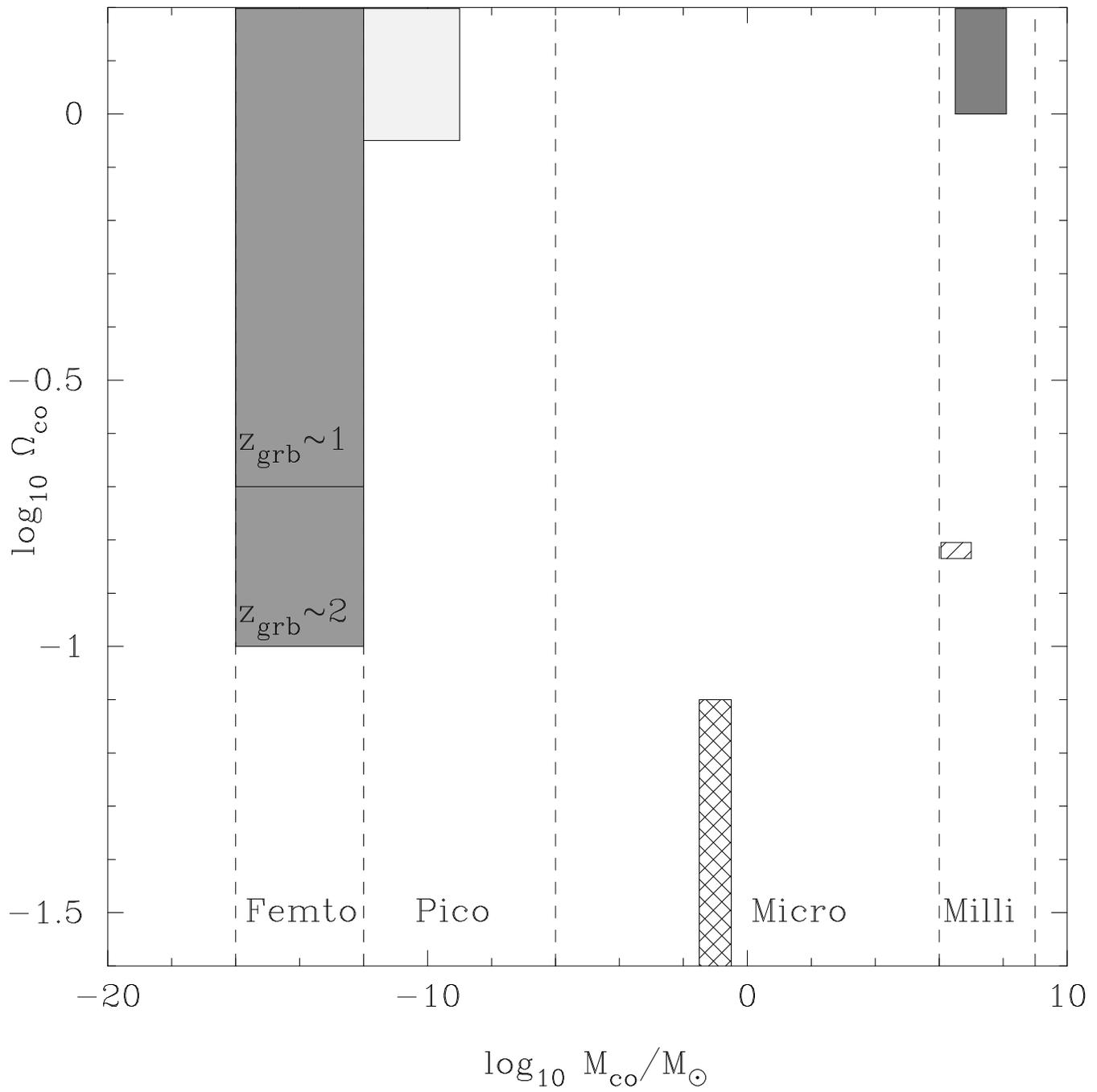
\end{document}